\documentclass[review,number,sort&compress]{elsarticle}

\usepackage{amssymb}
\usepackage{lineno}

\usepackage{graphicx}
\usepackage{subfig}
\usepackage{mathtools}
\usepackage{multirow}
\usepackage{tabularx}
\usepackage{textcomp}

\journal{Nuclear Instruments and Methods in Physics Research Section A}

\begin{document}

\begin{frontmatter}



\title{Design and construction of a multi-layer CsI(Tl) telescope for high-energy reaction studies}


\author[A1,A2]{D. Yan}
\author[A1]{Z.Y. Sun}
\author[A1]{K. Yue\corref{cor}}
\ead{yueke@impcas.ac.cn}
\cortext[cor]{Corresponding author. Tel.: +86 9314969321.}
\author[A1]{S.T. Wang}
\author[A1]{X.H. Zhang}
\author[A1]{Y.H. Yu}
\author[A1]{J.L. Chen}
\author[A1]{S.W. Tang}
\author[A1]{F. Fang}
\author[A1,A2,A3]{Y. Zhou}
\author[A1,A2,A3]{Y. Sun}
\author[A1,A2]{Z.M. Wang}
\author[A1,A2]{Y.Z. Sun}

\address[A1]{Institute of Modern Physics, Chinese Academy of Sciences, Lanzhou 730000, China}
\address[A2]{University of Chinese Academy of Sciences, Beijing 100049, China}
\address[A3]{Lanzhou University, Lanzhou 730000, China}

\begin{abstract}
A prototype of a new CsI(Tl) telescope, which will be used in the reaction studies of light isotopes with energy of several hundred AMeV, has been constructed and tested at the Institute of Modern Physics, Chinese Academy of Sciences. The telescope has a multi-layer structure and the range information will be obtained to improve the particle identification performance. This prototype has seven layers of different thickness. A 5.0\% (FWHM) energy resolution has been extracted for one of the layers in a beam test experiment. Obvious improvement for the identification of $^{14}$O and $^{15}$O isotopes was achieved by using the range information.
\end{abstract}


\end{frontmatter}


\section{Introduction}
\label{introduction}
Particle identification is one of the most essential tasks in nuclear physics experiments, especially for reaction studies. Various methods have been developed for this purpose, such as the $\Delta$$E-E$, the range ~\cite{Chulick1973,Greiner1972} as well as the Bragg peak amplitude ~\cite{Asselineau1982} and the time-of-flight (TOF) measurements ~\cite{Bass1975}. Every method has its advantages and limitations. For example, the energy loss ($\Delta$$E$) combined with the residual energy ($E$) measurement ($\Delta$$E-E$ method) can identify isotopes of light ions easily but fail for heavier ions due to the limited resolution of the detectors ~\cite{Carboni2012}.
It is often difficult to know a-priori the most efficient method for particle identification when designing a new experiment.

Normally, the atomic number $Z$ of a nucleus is easy to be determinated by the partial energy deposit ($\Delta$E) measurements in a thin detector, but the mass number $A$ is a bit troublesome, especially for high-energy cases.
Magnetic spectrometer combined with TOF measurements maybe a good solution, but it's too complicated for some simple measurements like charge changing cross-section (CCCS) or one nucleon knock-out.

A Calorimeter Telescope (CATE) was developed at GSI ~\cite{Wollersheim2005, Lozeva2006} and used in a series of high energy experiments to perform particle identification behind the reaction target ~\cite{Burger2005, Banu2005, Saito2008, Wieland2009} by using $\Delta$$E-E$ method. Though it met the demands in those experiments, the overlap between neighbouring isotopes in the identified mass spectrum is not negligible.

Simple calculations show that for high-energy particles with the same velocity, even neighbouring isotopes will have obvious difference in their ranges in materials. In this paper, a prototype of a multi-layer calorimeter telescope, which combines the $\Delta$$E-E$ and $\Delta$$E-Range$ methods, will be introduced for studying direct reactions, such as single nucleon knock-out. By combining the $\Delta$$E-E$ and $\Delta$$E-Range$ methods, decent PID performance has been achieved with this simple setup for light nuclei (Z$<$10) around 300 AMeV at the Cooler Storage Ring of the Heavy Ion Research Facility in Lanzhou (HIRFL-CSR)~\cite{Xia2002}.

\section{Design and manufacture of the new telescope}
\subsection{Design}
cIn order to stop the heavy ions with energy of several hundred AMeV, the detector material needs to have higher density, higher stopping power and large enough volume. We finally choose CsI(Tl), an inorganic scintillating material, as the detector material for this new telescope because except these requirements, it also has other benefits like relatively good energy resolution, easy for manufactory and lower price.

This new detector prototype was designed according to the range of oxygen isotopes with energy around 250 AMeV, which is about 36.6 mm for $^{16}$O from simple calculations.
The active area of the prototype is 7 $\times$ 7 cm$^2$, and the photomultiplier tubes (PMTs) would be used as readout device.
For the thickness of the layers, although smaller layer thickness can give good resolution for range, but it's not a good solution in reality. This is mainly because the light collection of a thin plate will strongly depend on the incident position, which may influence the energy resolution of the detector.
So we decided to have a prototype with 7 layers, and the thickness of these layers are 7 mm, 7 mm, 7 mm, 5 mm, 5 mm, 10 mm, and 10 mm in order.

\begin{figure}[!ht]
	\centering
	\subfloat{\includegraphics[scale=0.4]{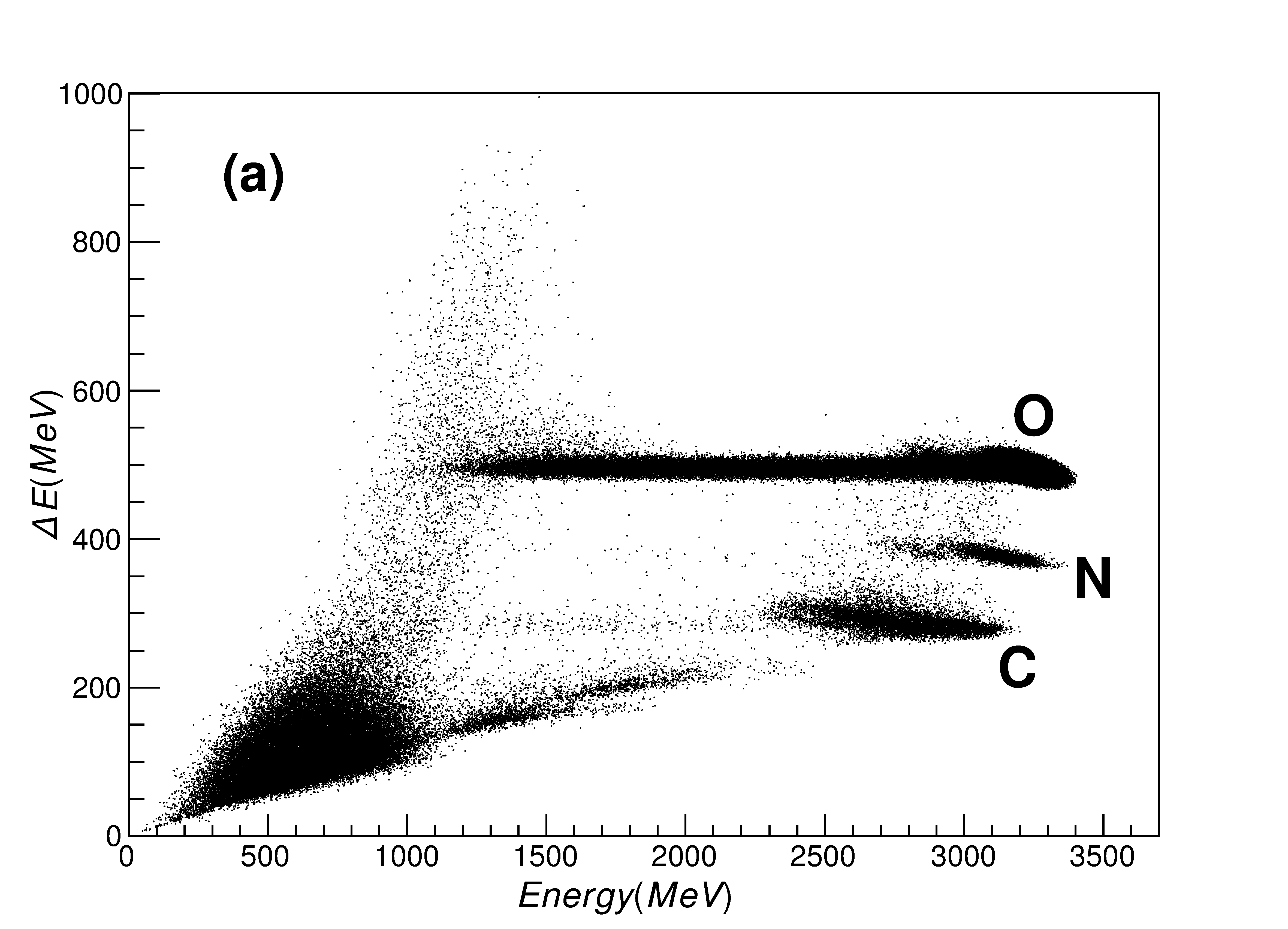} \label{subfig:Simu_DEvsE_WDR}} \\
	\subfloat{\includegraphics[scale=0.4]{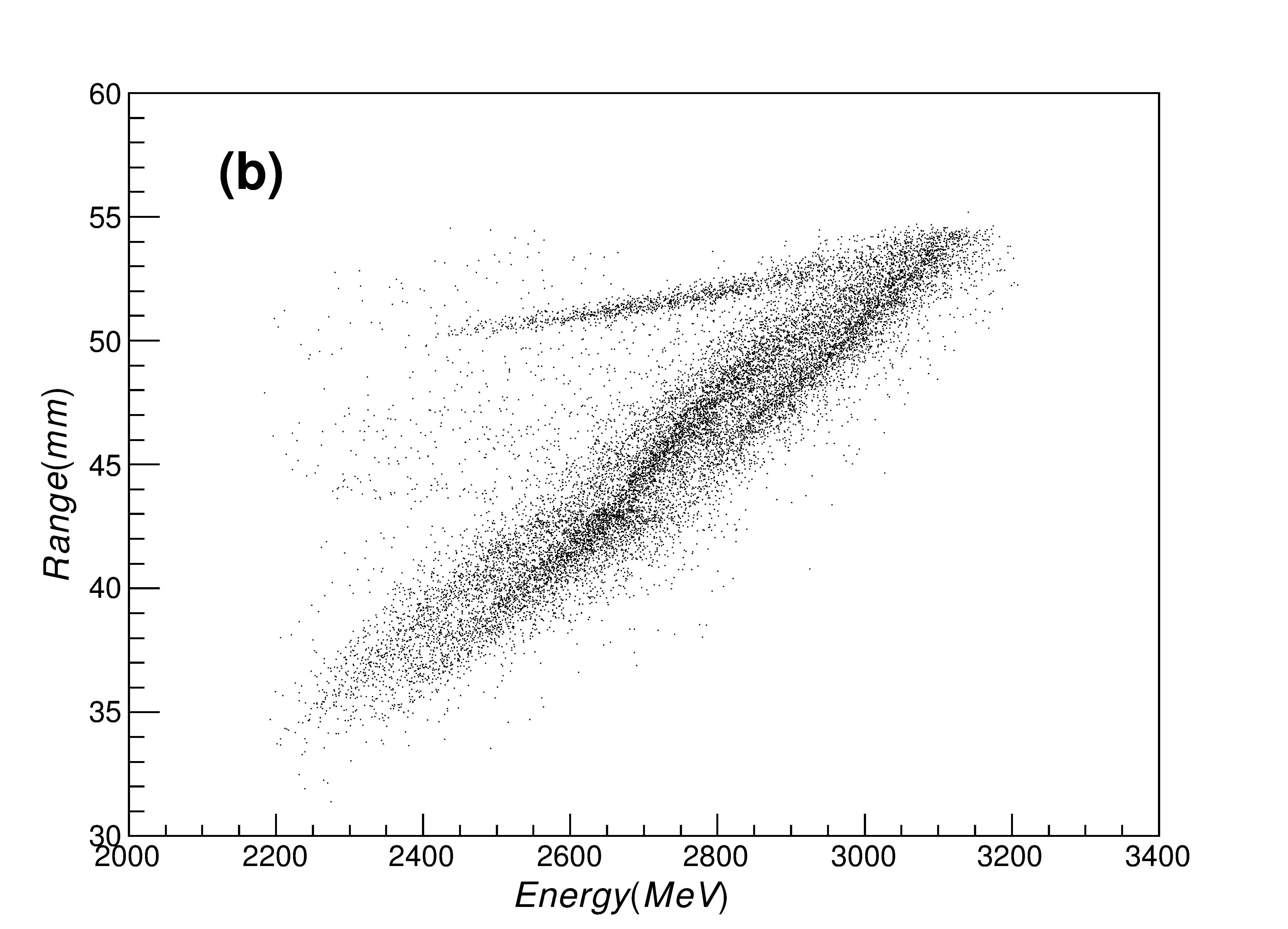} \label{subfig:Simu_DEvsE_WDR_R}}
	\caption{The PID spectra of the detector through Monte Carlo simulation with 2\% energy resolution for each layer of the telescope. (a) is the PID spectrum through $\Delta$$E-E$ method while (b) is the PID spectrum for Carbon isotopes through $Range-E$ method.} \label{fig:SimuPID}
\end{figure}

A Monte Carlo simulation based on Geant4~\cite{GEANT4} has been developed to evaluate the performance of the detector.
In the simulation, a 300 AMeV $^{14}$O beam hit a 6 g$/$cm$^2$ Al target and the products were detected by this telescope.
The obtained $\Delta$E-E spectrum is showing in Fig.\ref{subfig:Simu_DEvsE_WDR}, here a 2\% energy resolution was assumed for each detector layer and the first layer of the telescope was used as the $\Delta$E detector.
From the spectrum, we can see different elements clearly, but for the isotopes, the mixture between neighboured ones are serious and the identification are very difficult.
Fig.\ref{subfig:Simu_DEvsE_WDR_R} shows the Range-E spectrum for carbon isotopes selected in Fig.\ref{subfig:Simu_DEvsE_WDR}. In the spectrum, the range of one particle, which was stopped in layer n of the telescope, was simply calculated by
\begin{equation}
R=\sum_{i=1}^{n-1} D_{i} + \alpha\cdot\cfrac{\Delta E_{n}}{\Delta E_{n-1}/D_{n-1}} ,
\label{eq:Rn}
\end{equation}
here $\Delta E_{i}$ is the energy deposition in layer i and $D_{i}$ is the thickness of layer i. $\alpha$ is an adjustable parameter. 
In Fig.\ref{subfig:Simu_DEvsE_WDR_R}, we can separate different isotopes, such as $^{10}$C, $^{11}$C and $^{12}$C, easily, which means the particle identification ability will be improved a lot with the help of the range information, and this new telescope will be useful in high energy experiments.

\subsection{The manufacture}
All the CsI(Tl) crystals were produced and machined in the Institute of Modern Physics, Chinese Academy of Sciences (CAS)~\cite{Chen2008}, which have supplied this kind of crystals to many other projects~\cite{Yue2013} and the performance of the crystals are reliable.

The PMTs used for this prototype are R4443 from Hamamatsu, which is a 10-stage linear tube with low noise and an active area of 10 mm diameter.

For energy deposit measurements using scintillator detectors, the uniformity of light collection is very important especially for those who have large volume or special shape.
Because the thickness of one layer are much smaller compared with the other dimensions, the scintillation lights can only be collected from its edge side and that makes the light obtained with one PMT very sensitive to the hit position of the incident particle. For a long bar shape detector, this normally solved by read out from the both ends of the detector and use the $\sqrt{Q_{1} \times Q_{2}}$ method for energy reconstruction~\cite{Liu2004}.
But for a planar one like ours, this method is not suitable.
A more common way for solving this problem is to make position-dependent corrections for the light output, as Ref.~\cite{Wagner2001,Goethem2004} did.
The adverse aspect of this method is that additional position sensitive detectors must be needed together with your detector, which makes the system much more complicated especially for a multi-layer detector.

There are many reasons that may influence the uniformity of light collection, the number of PMTs used and the reflecting material are more important in our case.

To study this, we followed the method in Ref.~\cite{Wagner2001,Goethem2004}. The uniformity factor G is defined as
\begin{equation}
G=\cfrac{\left |Q_{a}-Q_{c} \right |}{Q_{c}} ,
\label{eq:G}
\end{equation}
here $Q_{c}$ is the light collected while the incident occured at the center of the layer and $Q_{a}$ is the light collected at an arbitrary position.
The layer studied was average divided into 9 parts, one at the center and the other eight surroundings like a pound sign.
We measured the light output while using a collimated $^{239}$Pu $\alpha$ source at the center of each part in vacuum, and got $Q_{c}$ and $Q_{a}$ by summing all the PMTs used. The G factor was calculated by averaging the results of the eight surrounding parts.

\begin{figure}[!ht]
	\centering
	\includegraphics[scale=0.06]{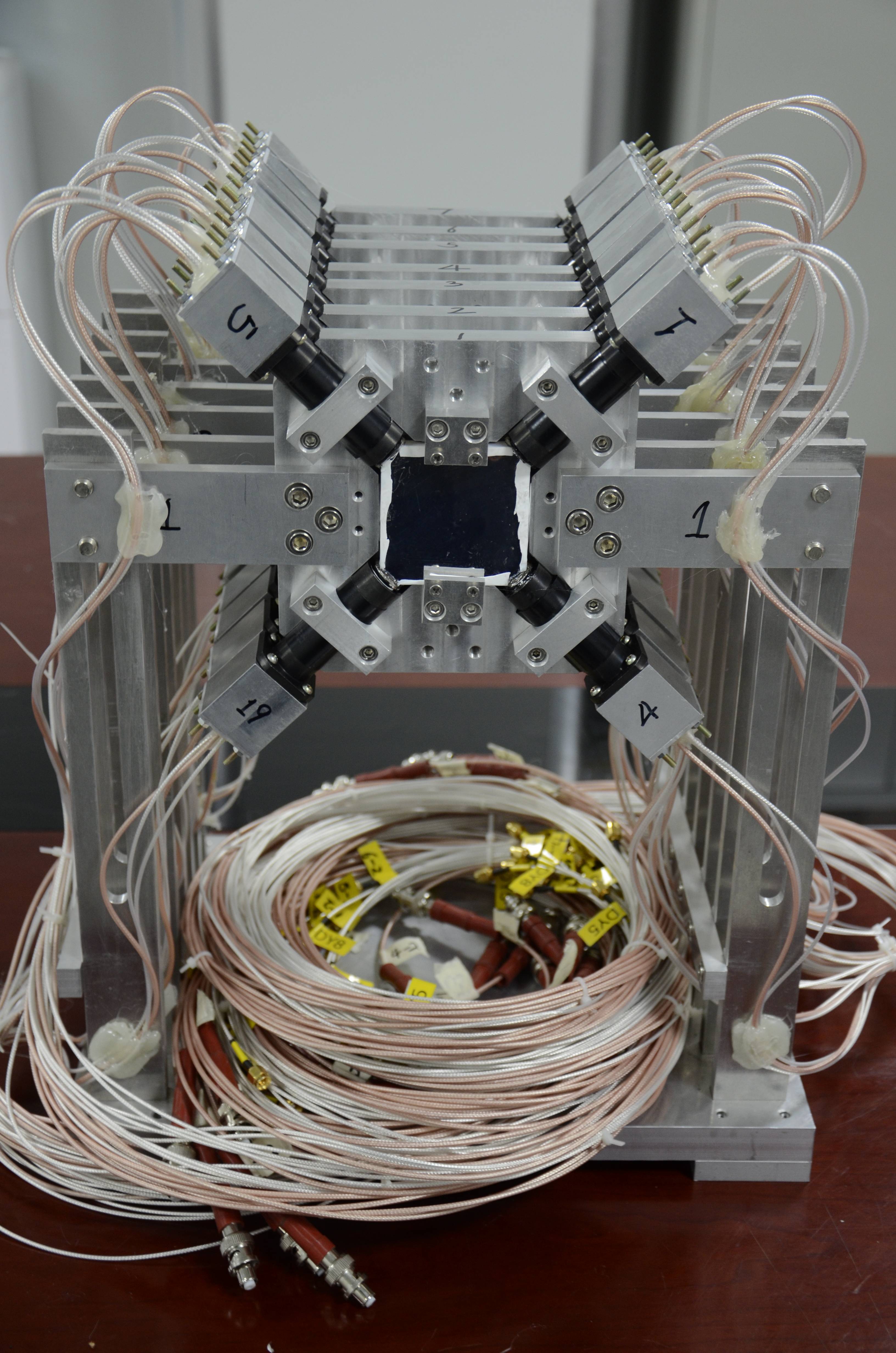} \label{fig:photo}
	\caption{Prototype of the multi-layer CsI(Tl) telescope.} \label{fig:Detector}	
\end{figure}

We found that the G factor become smaller as more PMTs used for light collection, for example, the G factor is 9.7\% while using 4 PMTs and 23.7\% for 2 PMTs. We also found that read out from the the corners give better uniformity than read from the center of the edge sides.
The prototype of the new telescope is shown in Fig. \ref{fig:Detector}. For every layer, we have 4 PMTs, and the corners of the crystal were truncated for 5 mm to couple with the PMTs.

Due to the relatively small thickness of our layer, only very few photons can reach the PMTs without any reflection by the refecting materials at the crystal surface. In each reflection, not all the photons would be reflected, and this kind of reflection may occurred many times, which is also influenced by the reflection pattern (mirror or diffused reflection) of the material. So the loss of photons would be a considerable number, and choosing a correct reflecting material is very important.

\begin{figure}[!ht]
	\centering
	\subfloat[ESR]{\includegraphics[scale=0.25]{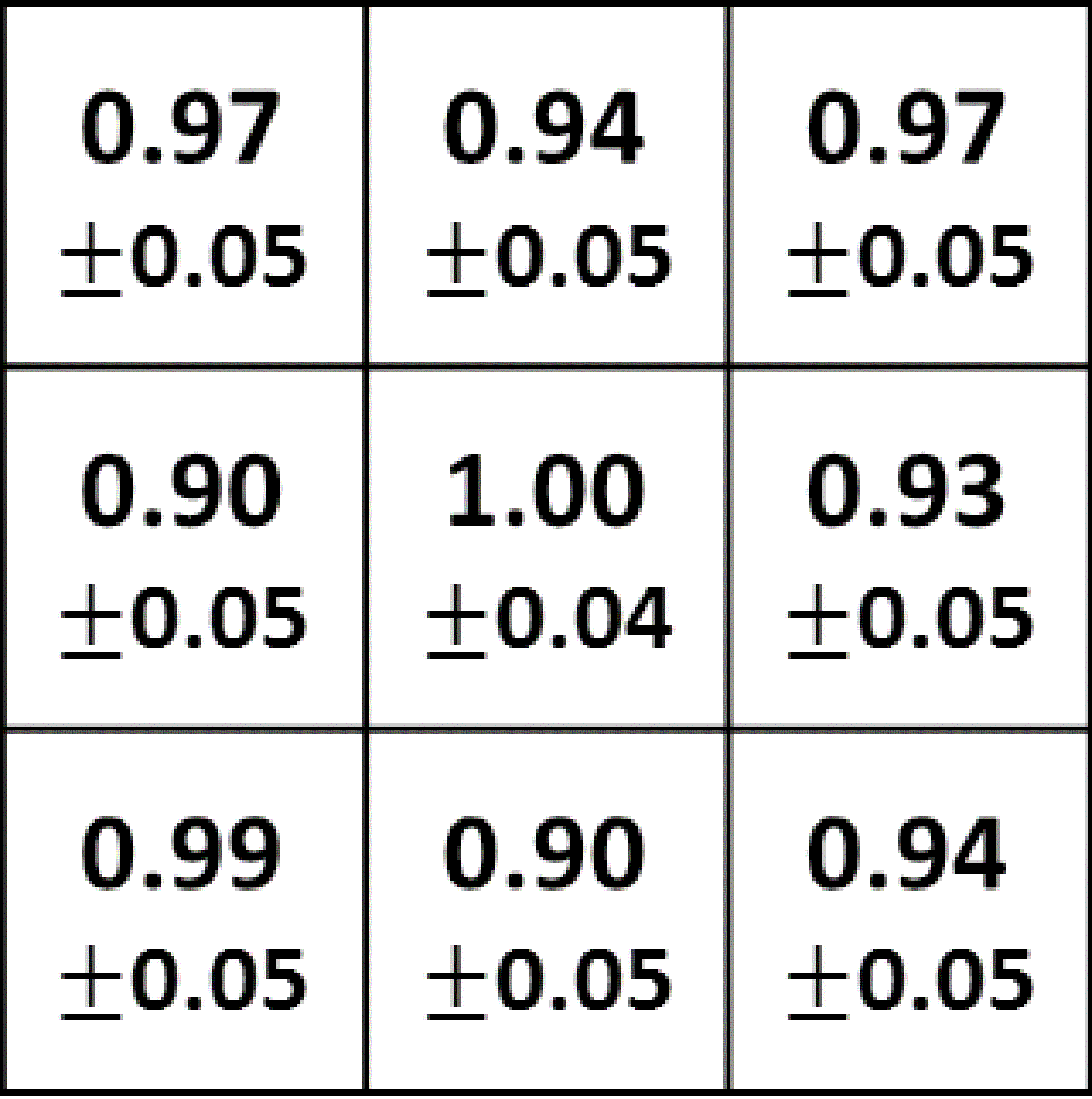} \label{subfig:ESR}} \hspace{20pt}
	\subfloat[aluminized Mylar]{\includegraphics[scale=0.25]{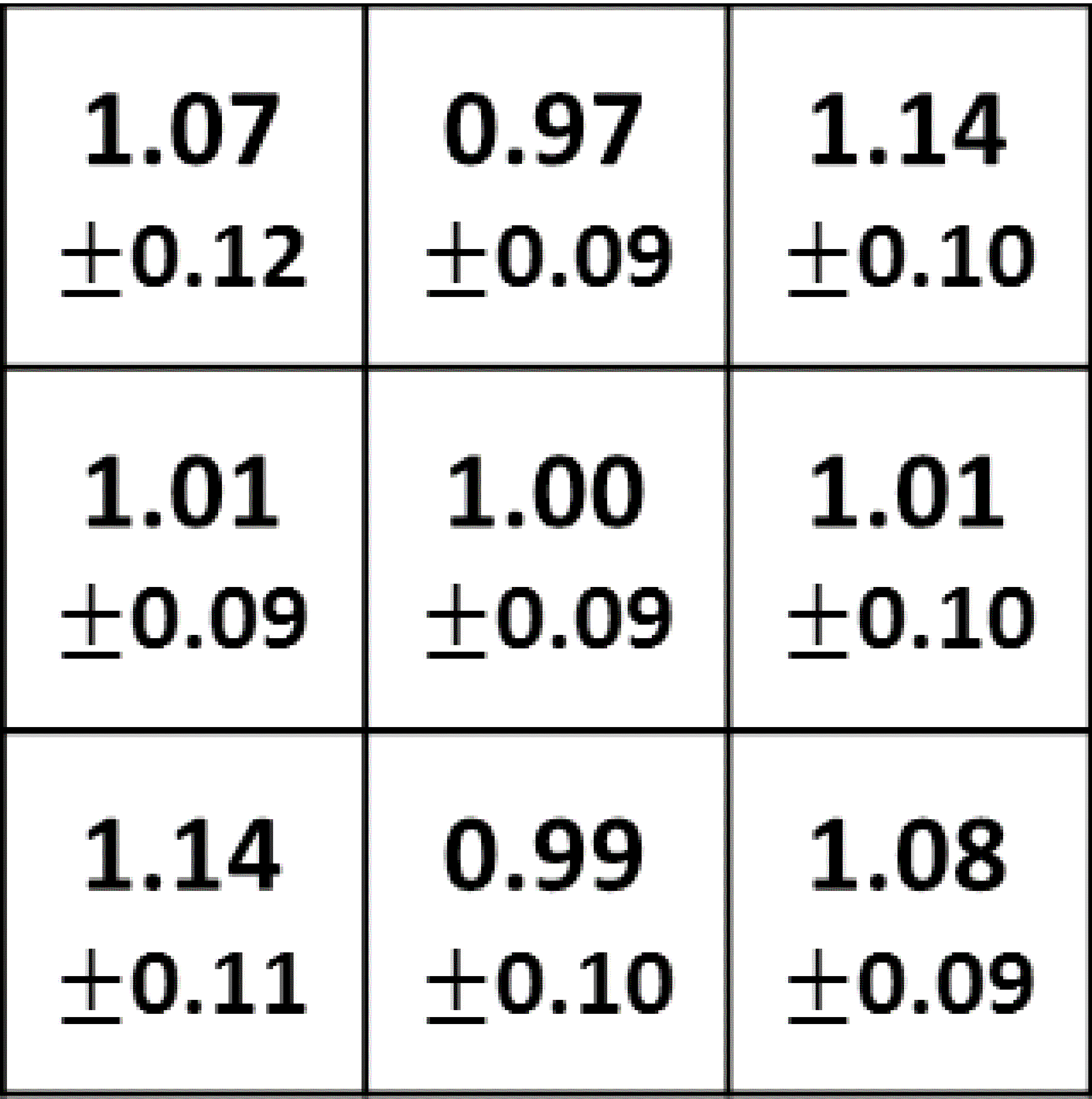} \label{subfig:Mylar}} \\
	\subfloat[Tyvek paper]{\includegraphics[scale=0.25]{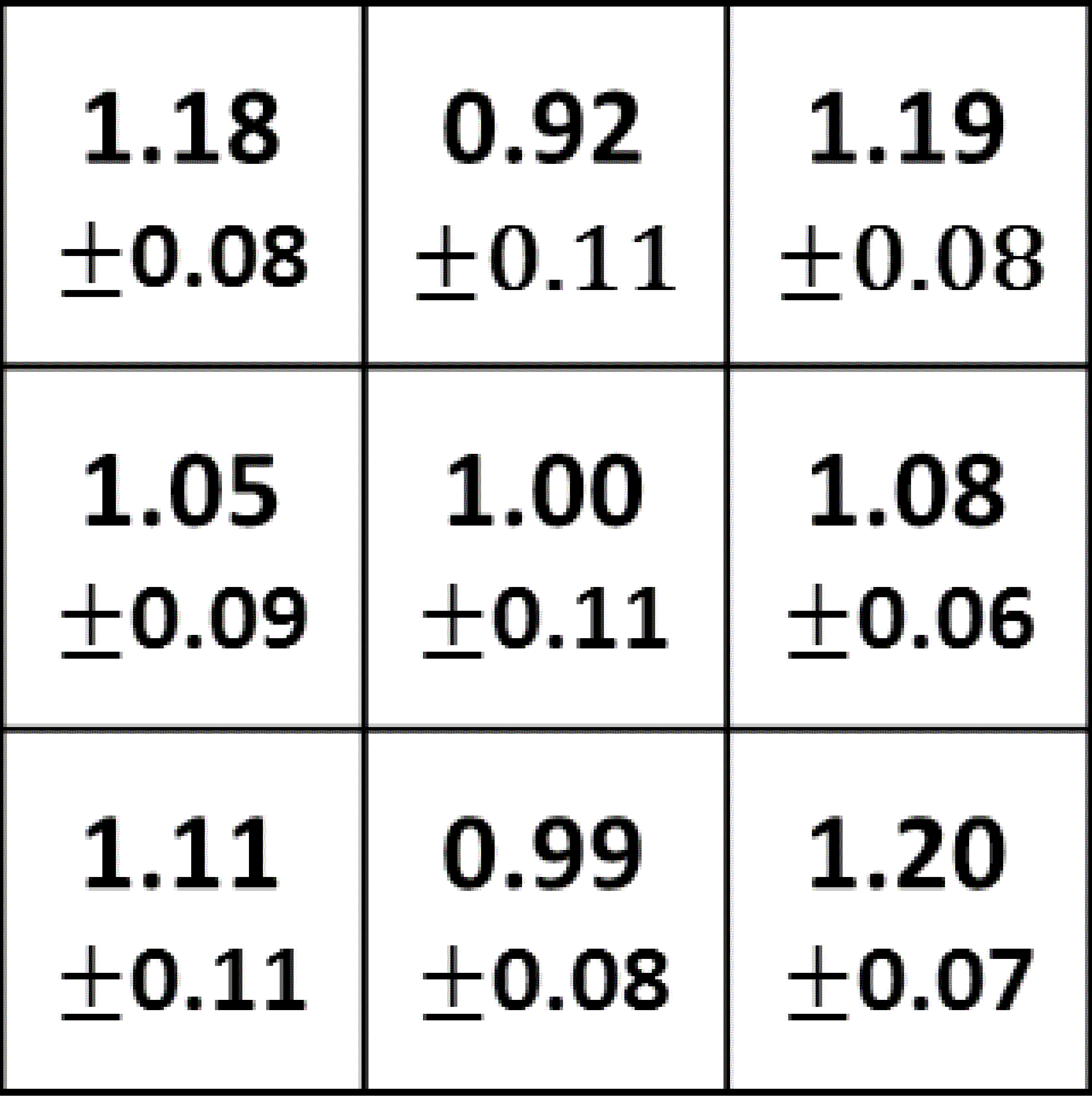} \label{subfig:Tyvek}} \hspace{20pt}
	\subfloat[aluminum foil]{\includegraphics[scale=0.25]{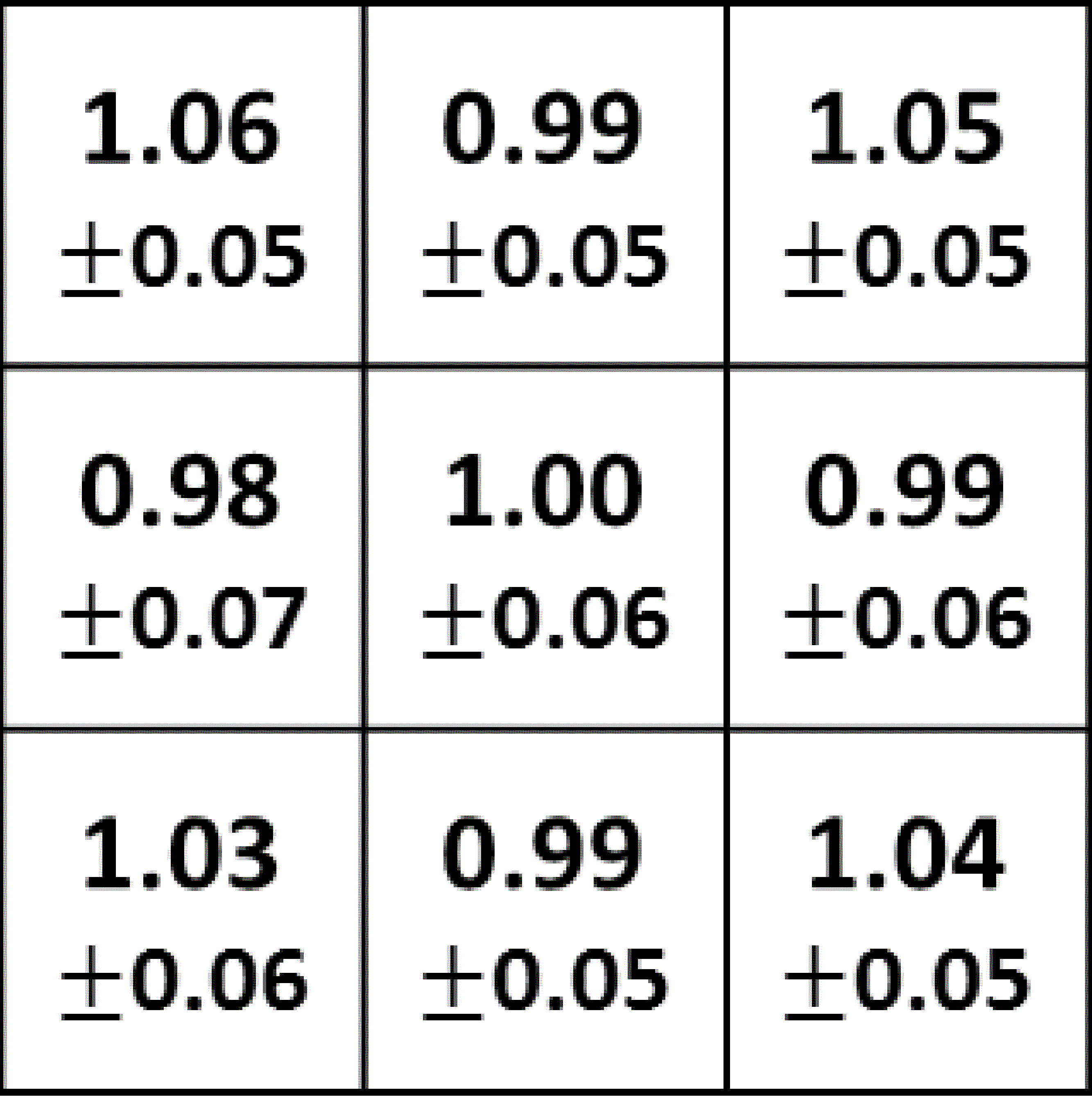} \label{subfig:Al}} \\
	\subfloat[aluminized Kapton]{\includegraphics[scale=0.25]{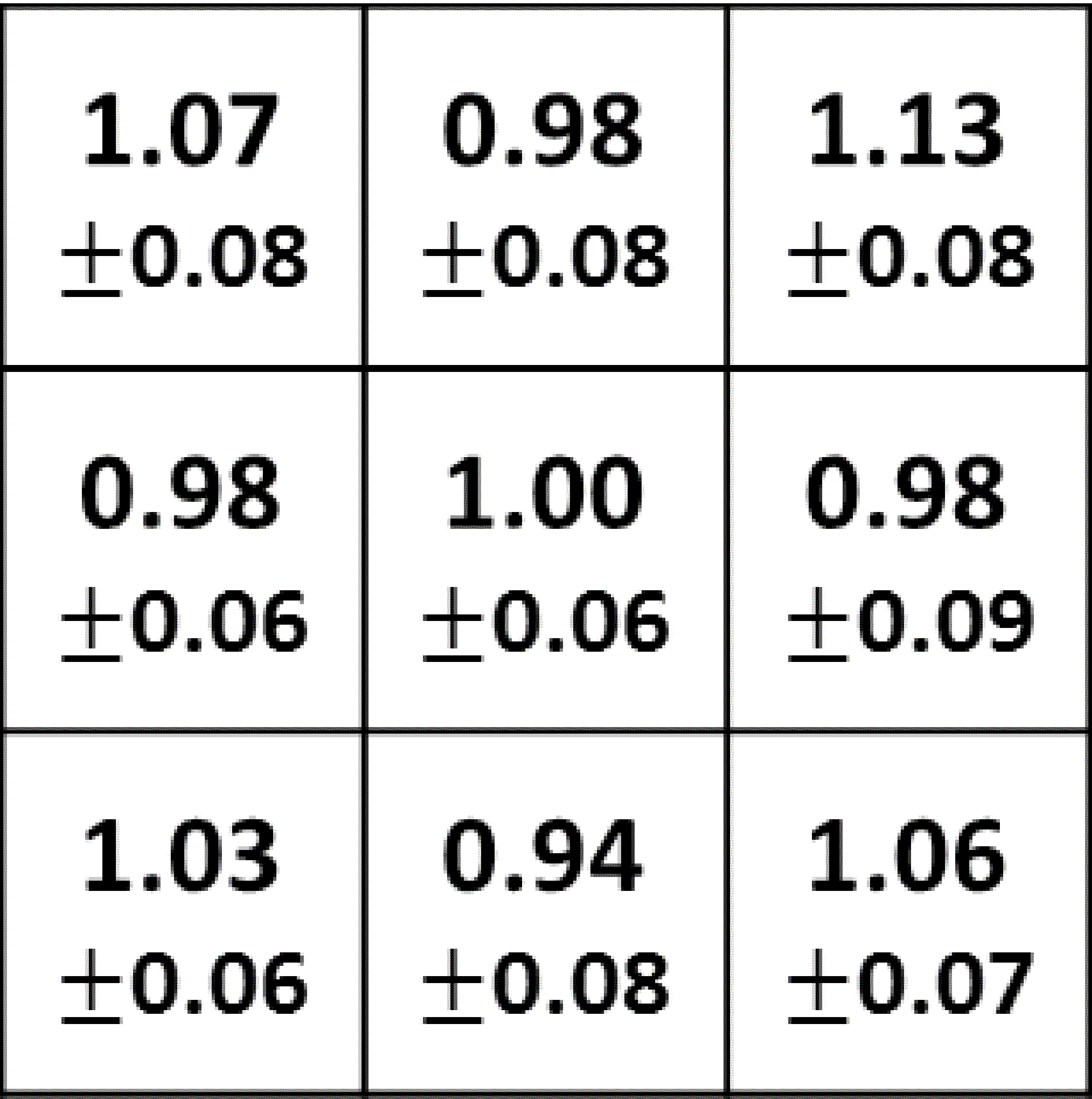} \label{subfig:Kapton}}
	\caption{The uniformity of CsI(Tl) crystal for different coating.} \label{fig:Uniform_Coating}
\end{figure}

Many different materials, such as Vikuiti\textsuperscript{TM} Enhanced Specular Reflector film (ESR) provided by 3M company ~\cite{3MCOMPANY}, aluminized Mylar foil, Tyvek paper, aluminum foil and aluminized Kapton foil were studied in this work, and the results are shown in Fig. \ref{fig:Uniform_Coating}. The corresponding G factors were also calculated and shown in Tab. \ref{tab:coating}.
Except the uniformity, the reflecting materials also have large influence on the energy revolution of the detector. The energy revolution for the $\alpha$ from $^{239}$Pu source impinging on the center of the layer is also listed in Table \ref{tab:coating} for different materials.

From these results, we can see that the ESR film has the best energy resolution while the aluminum foil has the best light collection uniformity. We prefer to use the ESR film as the reflecting material for our detector, but the thickness of this film is a little larger (~80 $\mu$m), may cause large error for nearly stopped incident ions. So we finally wrapped the first three layers with the ESR film, and the other four with the aluminum foil, which also has good energy resolution with a much smaller thickness of 7 $\mu$m.

\begin{table}[!ht]\centering
	\scriptsize
	\newcommand{\tabincell}[2]{\begin{tabular}{@{}#1@{}}#2\end{tabular}}	
	\caption{peak channal and energy resolution for each coating materials} \label{tab:coating}
	\renewcommand{\arraystretch}{1.2}
    \tabcolsep=8pt
	\begin{tabularx}{120mm}{c X X X X X}\hline
		coating materials & ESR &\tabincell{c}{aluminized \\ Mylar} & Tyvek & Aluminum & \tabincell{c}{aluminized \\Kapton} \\ \hline
		peak channal & 5850.7 & 2510.9 & 2924.7 & 4937.9 & 2820.7 \\
		\tabincell{c}{energy resolution \\(FWHM)} & 9.6\% & 20.9\% & 26\% & 14.6\% & 14.7\% \\
		G factor & 6.0\% & 6.1\% & 11.3\% & 2.9\% & 5.1\% \\ \hline
	\end{tabularx}
\end{table}

\subsection{The electronics and DAQ system}
Because the particles may stop inside the telescope, the energy deposit in each layer may differ from few MeV to some GeV, for example, the maximum energy loss of $^{16}$O in a 5 mm layer is about 1.2 GeV. It's difficult to get good energy resolution with this large dynamic range by a single electric circuit. To make a balance between the dynamic range and the energy resolution, we designed a special base circuit for the PMTs, which read the signals out from two different dynodes, the dynode 5 and dynode 8, which has a difference of about 50 times in signal amplitudes.

Consider that our prototype has 7 layers, which means 28 PMTs and 56 signal channels, an electronics system with high integration is needed. The one we chose is a data processing board ~\cite{Yang2015} developed for the DAMPE project ~\cite{Chang2014}, in which it couples with PMTs and scintillators also. This board is developed in the IMP, and it's base on the popular used ASIC chip VA32, which integrates the charge-sensitive preamplify, shaper and peak holding circuits for 32 channels in one chip, provided by the Integrated Detector Electronics AS (IDEAS) company in Norway ~\cite{IDEAS}.
Each board has two VA32 chips, which means 64 channels, coupled to a 14 bit Analog to Digital Converter (ADC) through a multiplex switcher. The work of the board is controlled by a onboard field-programmable gate array (FPGA) chip, and the data is temporary stored in the FPGA after conversion.

The DAQ system includes a specific designed board that can support up to 4 data processing boards, with the help of its onboard FPGA, the data from all the boards are collected, assembled and transferred to the computer through the USB interface. The associated DAQ software is developed using National Instruments VISA library ~\cite{VISA} in the LINUX system.

The stability and reliability of this system (both hardware and software) have been approved by the DAMPE project, and the dead time of the whole system (include one data processing board) is 280 $\mu$s.

\section{Performance of the telescope prototype}

Every layer for this prototype was checked with $^{239}$Pu $\alpha$ source after manufacture, and the energy resolution (FWHM) is better than 10\%, which corresponds to 500 KeV for 5 MeV $\alpha$ particle, and it seems like that an energy resolution of better than 2\% can be obtained for particles with energy deposition of several hundred MeV.

Because the energy of $\alpha$ particles from the source is too small compared with the region we are interested in, the results estimated from it maybe far from the reality as this new telescope is designed for high energy experiments. So we used the high energy beam to check whether this new detector can work well and benefit our requirements.

The beam test was performed at the external target experiment hall of CSRm. A primary $^{16}$O  beam with the energy of 360 AMeV bombarded a beryllium production target, and the produced secondary beam was separated and transferred to the hall through the fragment separator RIBLL2. The particle identification of incident beam was done by the TOF-B$\rho$-$\Delta$E method with the help of some detectors along the beam line, and the position and trajectory of incident particles at the target are also known.

\begin{figure}[!ht]
	\centering
	\includegraphics[scale=0.4]{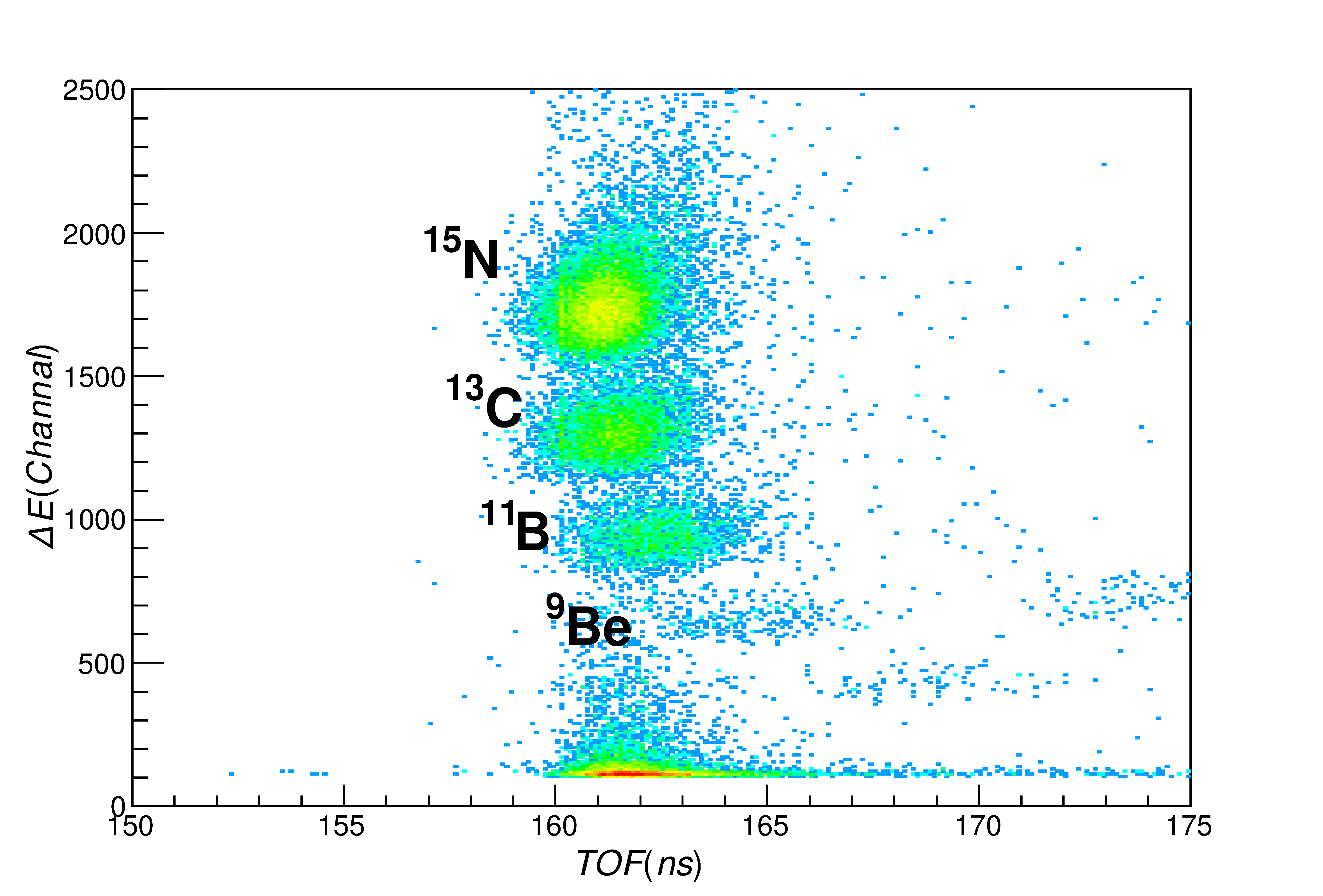}
	\caption{The spectrum of energy deposition against time of flight for the secondary beam.} \label{fig:Cali_SivsTOF}	
\end{figure}

Because the total energy is obtained by summing the energy loss information from all the layers, the layers need to be calibrated first. This was done by using a cocktail beam, which is a group of neutron-rich nuclei with similar A/Q, as shown in Fig. \ref{fig:Cali_SivsTOF}, and all the layers were calibrated separately.

For scintillators as the CsI(Tl), we know that the relation between the light output and the energy deposition is not linear, which means that except the energy deposition, the light output is related to the charge number $Z$ and mass number $A$ of the incident ions also.
These make the energy calibration for CsI(Tl) rather complicated, and we can not explain this analytically till now. Many works had been done in this field by introducing empirical functions to fit the measured data ~\cite{Womack1966, Twenhofel1990, Valtonen1990, Colonna1992, Horn1992, Mastinu1994, Larochelle1994}.

\begin{figure}[!ht]\centering
	\includegraphics[scale=0.4]{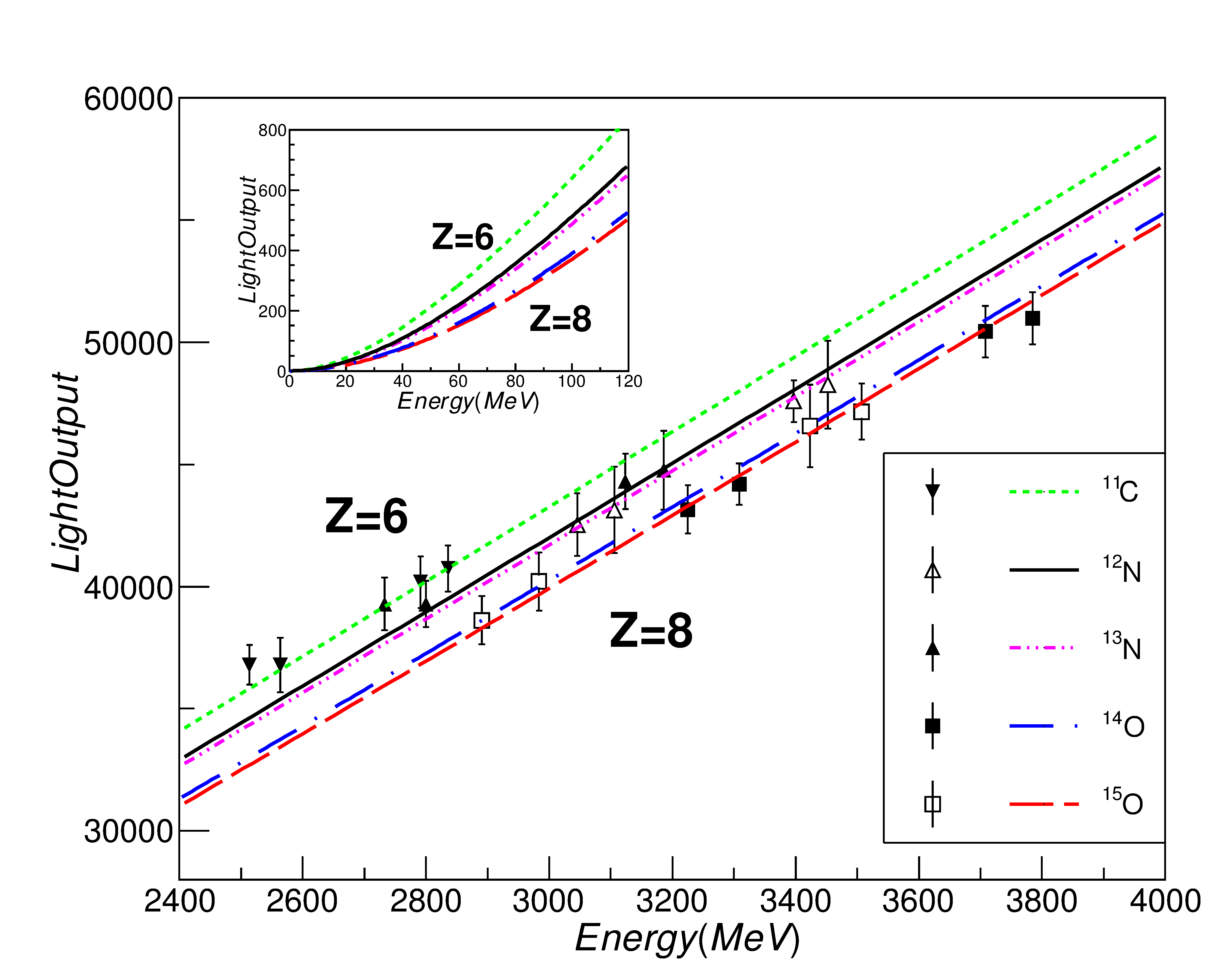}
	\caption{Total light output against the energy for different incident particles in all of the crystals. The symbols are experimental data. The different curves are the result of a least-squares fit to Eq.(9) in Ref ~\cite{Parlog2002_2}. The inset picture shows the curves in the low energy region.}
	\label{fig:LvsE_all}
\end{figure}

In our test, we put some aluminum made degraders with different thickness (22.5mm, 21mm, 13.5mm, 12mm) upstream of the telescope to get more different energy points, and selected the same function as M. P$\hat{a}$rlog ~\cite{Parlog2002_1, Parlog2002_2} for the energy calibration. Fig. \ref{fig:LvsE_all} shows the calibration results for the telescope, and detailed description can be found in ref.~\cite{Yan2015}.

\begin{figure}[!ht]
	\centering
    \includegraphics[scale=0.4]{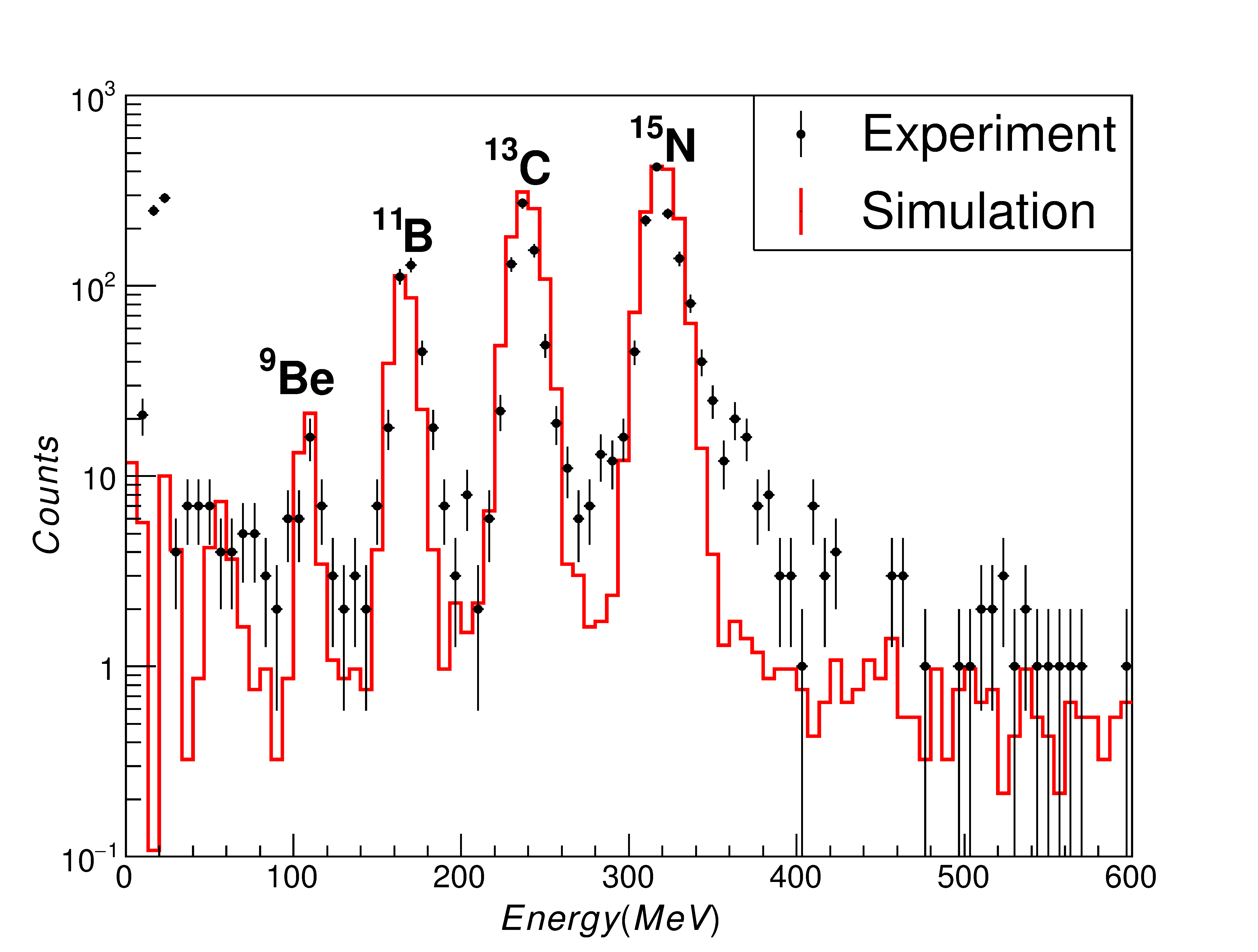}
	\caption{The spectra of energy deposition in a 7 mm thick detector layer. The black points are the beam test results and the red solid line is correspond simulation.} \label{fig:Cali}
\end{figure}

The beam spot is defocused to about 40 $\times$ 40 mm$^2$ to cover a large part of the telescope in this test, and Fig. \ref{fig:Cali} shows the calibrated energy loss spectra of the incident cocktail beam in a 7mm thick layer. Considering the energy spread of the incident beam and comparing the obtained spectrum with the Monte Carlo simulation results, an energy resolution of 5\% (FWHM) for 300 AMeV $^{15}$N, which deposited around 329 MeV energy in the crystal, is deduced.

This is worse than we expected, one important reason is the light collection uniformity. We selected the events with a much smaller beam spot, which was 10 $\times$ 10 mm$^2$, by using the incident trajectory information, the energy resolution changed from 7\% to 5\% (just a fit from the obtained energy loss spectrum, nothing deducted). This means the dependence of light collection and the incident position is still non-ignorable, and a position correction is needed to get a better energy resolution. Another reason is the abnormal larger noise in the electronics. Our system was disturbed by some unknown source suddenly appeared in the test, which caused the noise level to be several times larger than normal, and we had no chance to find them due to the limited beam time.

\begin{figure}[!ht]
	\centering
	\subfloat{\includegraphics[scale=0.4]{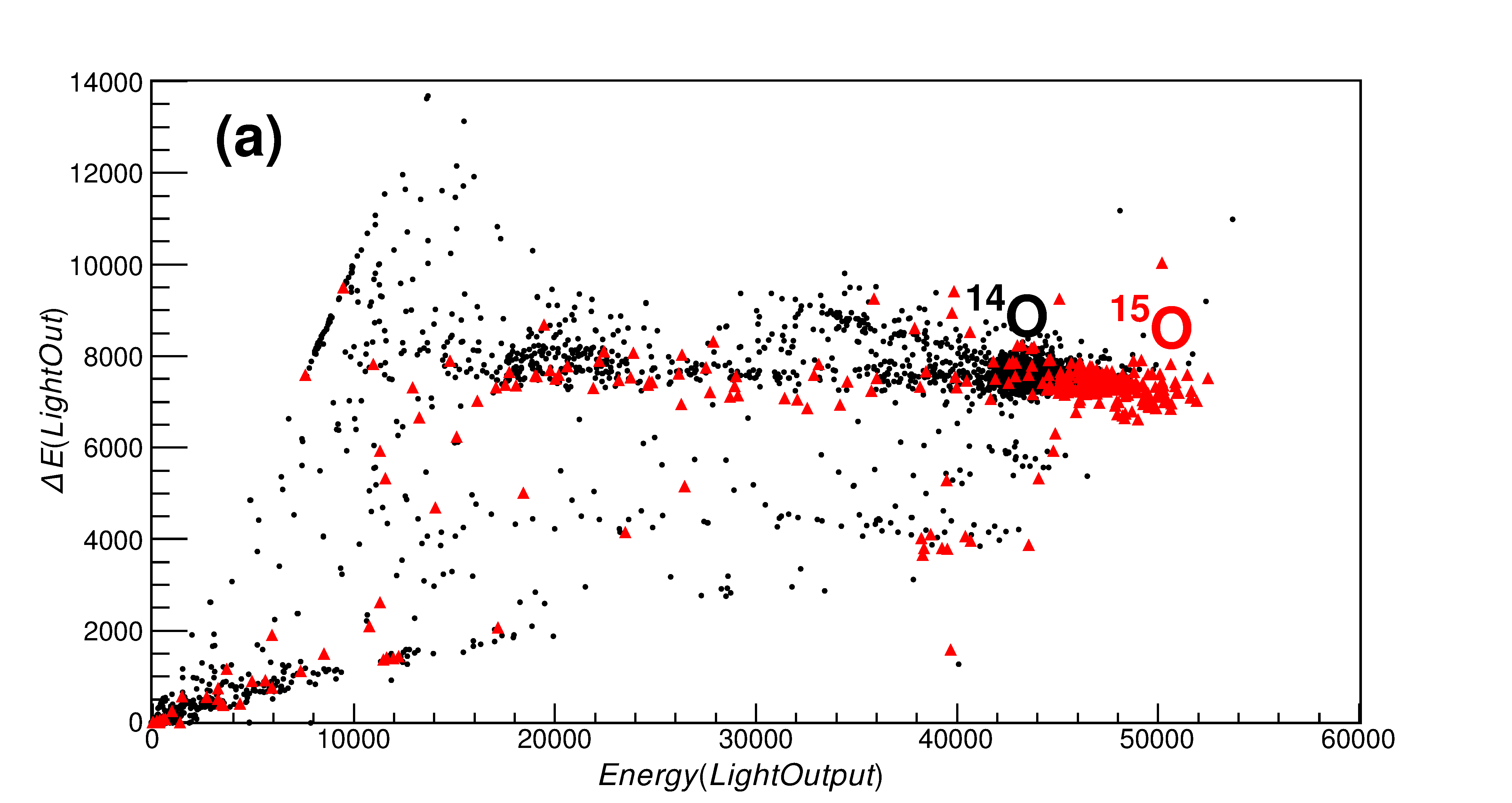} \label{subfig:DeltaE_E_D}}\\
	\subfloat{\includegraphics[scale=0.4]{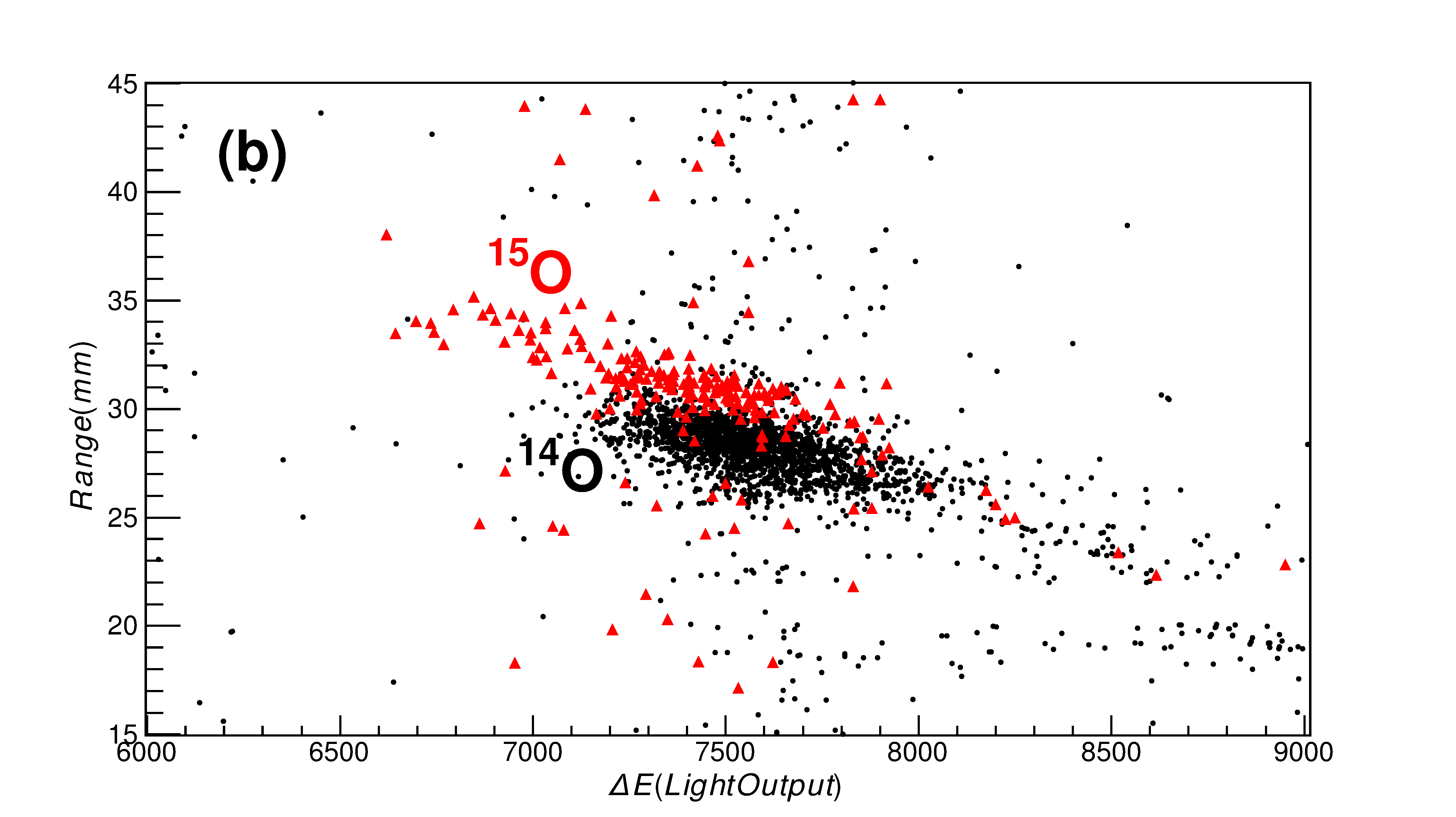} \label{subfig:Range_E_D}}
	\caption{The PID spectrum for Oxygen Isotopes in the detector. (a) is the spectrum through $\Delta$$E-E$ method while (b) is the spectrum through $Range-$$\Delta$$E$ method.} \label{fig:exp}
	
\end{figure}

To test the performance of the whole telescope, a cocktail beam included both $^{14}$O and $^{15}$O was used, and the particle identification of the incident beam was done by the detectors alone the RIBLL2.
An aluminum made degrader with the thickness of 22.5 mm was inserted as a target just up-stream of the telescope.
The first layer of the telescope was used as the $\Delta$E detector in the test, and Fig. \ref{subfig:DeltaE_E_D} shows the $\Delta$E-E spectrum obtained by selecting incident $^{14}$O (black dots) and $^{15}$O (red triangles) beams separately.

Due to the poor statistics, it's impossible to study the reaction products in this simple test, so we just studied the behavior of $^{14}$O and $^{15}$O beams, after penetrating the target, in the detector.
From Fig.\ref{subfig:DeltaE_E_D}, a clear separation between $^{14}$O and $^{15}$O is very difficult because they have a large overlap in total energy, and their difference in $\Delta$E is smaller due to the large velocity. But if we calculated the range as Eq.\ref{eq:Rn}, we can see a relatively much clear separation from the range vs $\Delta$E spectrum shown in Fig.\ref{subfig:Range_E_D}.
We think this means the idea of the telescope works well, although the results of our test run is not good enough due to the poor resolution of the detector prototype.

\section{Summary}
The prototype of a new CsI(Tl) telescope has been constructed and tested in the Institute of Modern Physics, CAS.
The idea of this telescope is to have a multi-layer structure, which can offer the range information except the normal $\Delta$E and E signals. And this range information is very helpful in getting better isotope identification for light isotopes with energy of several hundred AMeV.
To check this idea, we made a detector prototype with 7 layers in different thickness. And for one layer, we get a 5.0\% (FWHM) energy resolution with 330 MeV energy deposition from beam test. Even in this resolution, we can still get large improvement in separating the $^{14}$O and $^{15}$O isotopes with the help of their range information.
This indicates that the idea works well, and with improvement for the performance of the detector in the future, it will be a powerful tool in our future works.

\section*{Acknowledgement}
This work was supported by the National Natural Science Foundation of China (Grants No.11205206, No.11305222, No.11405242, No.11222550 and No.U1332207), and we also wish to express our gratitude to the accelerator crew at HIRFL for providing the beams.




\section*{References}
\begin{thebibliography}{99}


\bibitem{Chulick1973}E. Chulick, {\it et al.}, Nucl. Instr. and Meth. {\bf 109} (1973), 171.
\bibitem{Greiner1972}D.E. Greiner, {\it et al.}, Nucl. Instr. and Meth. {\bf 102} (1972), 291.
\bibitem{Asselineau1982}J. Asselineau, {\it et al.}, Nucl. Instr. and Meth. {\bf 204} (1982), 109.
\bibitem{Bass1975}R. Bass, {\it et al.}, Nucl. Instr. and Meth. {\bf 130} (1975), 125.
\bibitem{Carboni2012}S. Carboni, {\it et al.}, Nucl. Instr. and Meth. A {\bf 664} (2012), 251.
\bibitem{Wollersheim2005}H.J. Wollersheim, {\it et al.}, Nucl. Instr. and Meth. A {\bf 537} (2005), 637.
\bibitem{Lozeva2006}R. Lozeva, {\it et al.}, Nucl. Instr. and Meth. A {\bf 562} (2006), 298.
\bibitem{Burger2005}A. B\"{u}rger, {\it et al.}, Phys. Lett. B {\bf 622} (2005), 29.
\bibitem{Banu2005}A. Banu, {\it et al.}, Phys. Rev. C {\bf 72} (2005), 061305(R).
\bibitem{Saito2008}T.R. Saito, {\it et al.}, Phys. Lett. B {\bf 669} (2008), 19.
\bibitem{Wieland2009}O. Wieland, {\it et al.}, Phys. Rev. Lett. {\bf 102} (2009), 092502.
\bibitem{Xia2002}J.W. Xia, {\it et al.}, Nucl. Instr. and Meth. A {\bf 488} (2002), 11.
\bibitem{GEANT4}http://geant4.cern.ch/.
\bibitem{Chen2008}R.F. Chen, {\it et al.}, Chin. Phys C, {\bf 32} (02) (2008), 135.
\bibitem{Yue2013}K. Yue, {\it et al.}, Nucl. Instr. and Meth. B, {\bf 317} (2013), 653.
\bibitem{Liu2004}Liu Da-Zhi, {\it et al.}, High Energy Physics and Nuclear Physics {\bf 28} (2004), 186.
\bibitem{Wagner2001}A. Wagner, {\it et al.}, Nucl. Instr. and Meth. A {\bf 456} (2001), 290.
\bibitem{Goethem2004}M.-J. van Goethem, {\it et al.}, Nucl. Instr. and Meth. A {\bf 526} (2004), 455.
\bibitem{3MCOMPANY}http://www.3m.com/3M/en\_US/company-us/.
\bibitem{Yang2015}H.B. Yang, Ph.D. Thesis, University of Chinese Academy of Sciences (2015).
\bibitem{Chang2014}J. Chang, Chinese Journal of Space Science, {\bf 34} (2014), 550.
\bibitem{IDEAS}http://ideas.no/.
\bibitem{VISA}https://www.ni.com/visa/.
\bibitem{Womack1966}E.A. Womack, {\it et al.}, Phys. Rev. {\bf 144} (1966), 231.
\bibitem{Twenhofel1990}C.J.W. Twenh\"{o}fel, {\it et al.}, Nucl. Instr. and Meth. B {\bf 51} (1990), 58.
\bibitem{Valtonen1990}Eino Valtonen, {\it et al.}, Nucl. Instr. and Meth. A {\bf 286} (1990), 169.
\bibitem{Colonna1992}N. Colonna, {\it et al.}, Nucl. Instr. and Meth. A {\bf 321} (1992), 529.
\bibitem{Horn1992}D. Horn, {\it et al.}, Nucl. Instr. and Meth. A {\bf 320} (1992), 273.
\bibitem{Mastinu1994}P.F. Mastinu, {\it et al.}, Nucl. Instr. and Meth. A {\bf 338} (1994), 419.
\bibitem{Larochelle1994}Y. Larochelle, {\it et al.}, Nucl. Instr. and Meth. A {\bf 348} (1994), 167.
\bibitem{Parlog2002_1}M. P$\hat{a}$rlog, {\it et al.}, Nucl. Instr. and Meth. A {\bf 482} (2002), 674.
\bibitem{Parlog2002_2}M. P$\hat{a}$rlog, {\it et al.}, Nucl. Instr. and Meth. A {\bf 482} (2002), 693.
\bibitem{Yan2015}D. Yan, Ph.D. Thesis, University of Chinese Academy of Sciences (2015).


\end{thebibliography}


\end{document}